\documentclass[conference,a4paper]{IEEEtran}

\usepackage[utf8]{inputenc} 
\usepackage[T1]{fontenc}
\usepackage{url}
\usepackage{ifthen}
\usepackage{cite}
\usepackage[cmex10]{amsmath} 
\usepackage{amsfonts}
\usepackage[dvipdfmx]{graphicx,xcolor}

\begin{document}
\title{Cumulant Generating Function of Codeword Lengths in Variable-Length Lossy Compression Allowing Positive Excess Distortion Probability} 


\author{\IEEEauthorblockN{Shota Saito and Toshiyasu Matsushima}
\IEEEauthorblockA{Department of Pure and Applied Mathematics, Waseda University\\
3-4-1 Okubo, Shinjuku-ku, Tokyo, 169-8555 JAPAN \\ 
E-mail: wa-shota0425@fuji.waseda.jp and toshimat@waseda.jp}
}


\maketitle

\begin{abstract}
  This paper considers the problem of variable-length lossy source coding.
The performance criteria are the excess distortion probability and the cumulant generating function of codeword lengths.
We derive a non-asymptotic fundamental limit of the cumulant generating function of codeword lengths allowing positive excess distortion probability.
It is shown that the achievability and converse bounds are characterized by the R\'enyi entropy-based quantity.
In the proof of the achievability result, the explicit code construction is provided.
Further, we investigate an asymptotic single-letter characterization of the fundamental limit
for a stationary memoryless source.
\end{abstract}


\section{Introduction}
The problem of variable-length source coding is one of the fundamental research topics in Shannon theory.
For this problem, one of the criteria is the {\it normalized cumulant generating function of codeword lengths}.
This criterion was first proposed by Campbell \cite{Campbell} as a proxy for the mean codeword length.

Several previous works investigated the fundamental limit of the normalized cumulant generating function of codeword lengths:
e.g., 
\cite{Campbell} and \cite{Courtade1}
for the problem of variable-length lossless source coding;
\cite{Kuzuoka}
for the problem of variable-length source coding allowing errors;
\cite{Courtade2}
for the problem of variable-length lossy source coding.

The most relevant study to this paper is the work by Courtade and Verd\'u \cite{Courtade2}.
As described above, they considered the problem of variable-length lossy source coding.
As a criterion of the distortion measure,
they treated the {\it excess distortion probability}.
Their object of study was the code whose excess distortion probability is zero at a given distortion level $D$.
By using the {\it $D$-tilted R\'enyi entropy},
the study \cite{Courtade2} derived the converse bound for the fundamental limit of the normalized cumulant generating function of codeword lengths.

This paper considers the problem of variable-length lossy source coding and treats the same criteria as in \cite{Courtade2}.
However, the primary differences are 
1) we evaluate the code whose excess distortion probability may be  positive,
and
2) we derive both achievability and converse bounds by using a novel {\it R\'enyi entropy-based quantity}.
To show the achievability results, we give an explicit code construction instead of using the random coding argument.

Section II formulates the problem setup.
Section III describes the related work by Courtade and Verd\'u \cite{Courtade2}.
Sections IV and V show the main results in this paper.
In Section IV, we first define a R\'enyi entropy-based quantity.
Then, using this quantity,
we show non-asymptotic upper and lower bounds
of the fundamental limit.
Section V investigates an asymptotic single-letter characterization of the fundamental limit
for a stationary memoryless source.
Proofs of main results are in Section VI.
Section VII discusses the obtained results.

\section{Problem Formulation}

\newtheorem{theorem}{Theorem}
\newtheorem{condi}{Condition}
\newtheorem{defi}{Definition}
\newtheorem{lem}{Lemma}
\newtheorem{cor}{Corollary}
\newtheorem{proof}{Proof}
\newtheorem{remark}{Remark}
\newcommand{\argmax}{\mathop{\rm arg~max}\limits}
\newcommand{\argmin}{\mathop{\rm arg~min}\limits}

Let ${\cal X}$ be a source alphabet and ${\cal Y}$ be a reproduction alphabet, where both are finite sets.
Let $X$ be a random variable 
taking a value in ${\cal X}$ and $x$ be a realization of $X$.
The probability distribution of $X$ is denoted as $P_{X}$.
A distortion measure $d$ is defined as
$
d : {\cal X} \times {\cal Y} \rightarrow [0, +\infty).
$

The pair of an encoder and a decoder $(f, g)$ is defined as follows.
An encoder $f$  is defined as
$
f : {\cal X} \rightarrow \{ 0,1 \}^{\star},
$
where $\{ 0,1 \}^{\star}$ denotes the set of all finite-length binary strings 
and the empty string $\lambda$, i.e., 
$
\{ 0,1 \}^{\star} = \{\lambda, 0, 1, 00, \ldots \}.
$
An encoder $f$ is possibly {\it stochastic} and produces a non-prefix code.
For $x \in {\cal X}$, the codeword length of $f(x)$ is denoted as $\ell(f(x))$.
A {\it deterministic} decoder $g$ is defined as
$
g :  \{ 0,1 \}^{\star}  \rightarrow {\cal Y}.
$
Variable-length lossy source coding {\it without} the prefix condition  is discussed as in, for example,
\cite{Courtade2} and \cite{Kostina15}.
Once we prove a result for a non-prefix code,
we can easily derive a result for a prefix code.
We shall discuss it in Section \ref{Discuss}.

For a code $(f, g)$, we define the {\it excess distortion probability} and the {\it normalized cumulant generating function of codeword lengths}.

\begin{defi}
Given $D \geq 0$, 
the excess distortion probability is defined as
$
\mathbb{P} [ d(X, g (f (X))) > D ].
$
\end{defi}

\begin{defi}
Given $t > 0$, 
the normalized cumulant generating function of codeword lengths is defined as\footnote{All logarithms are of base 2 throughout this paper. Further, $\exp \{ \cdot \}$ denotes $2^{(\cdot)}$ in this paper.}
\begin{align}
\frac{1}{t} \log \mathbb{E} [2^{t \ell(f(X))}].
\end{align}
\end{defi}

\begin{remark} \label{cumulant}
The l'H\^{o}spital theorem yields
\begin{align}
\lim_{t \to 0} \frac{1}{t} \log \mathbb{E} [2^{t \ell(f(X))}]  & = \mathbb{E} [\ell(f(X))],  \label{m1} \\
\lim_{t \to \infty} \frac{1}{t} \log \mathbb{E} [2^{t \ell(f(X))}]  & = \max_{x \in {\cal X}} \ell(f(x)).
\end{align}
Thus, the normalized cumulant generating function of codeword lengths contains the mean codeword length and the maximum codeword length as its special cases.
\end{remark}

Using these criteria, we define a $(D, R, \epsilon, t)$ code.
\begin{defi}
Given $D, R \geq 0$, $\epsilon \in [0,1)$, and $t > 0$, 
a code $(f, g)$ satisfying
\begin{align}
\mathbb{P} [ d(X, g (f (X))) > D ] &\leq \epsilon \label{NP1}, \\
\frac{1}{t} \log \mathbb{E} [2^{t \ell(f(X))}]  &\leq R \label{NP2}
\end{align}
is called a $(D, R, \epsilon, t)$ code.
\end{defi}

The fundamental limit that we investigate is 
\begin{align}
R^{*} (D, \epsilon, t) & := \inf \{ R : \mbox{$\exists$ {\rm a} $(D, R, \epsilon, t)$ {\rm code}} \}. \label{Fundamental}
\end{align}

When we work on the setup of blocklength $n$,
we formulate the problem as follows.
Let ${\cal X}^n$ and ${\cal Y}^n$ be the $n$-th Cartesian product of ${\cal X}$ and ${\cal Y}$, respectively.
Let $X^n$ be a random variable 
taking a value in ${\cal X}^n$ and $x^n$ be a realization of $X^n$.
The probability distribution of $X^n$ is denoted as $P_{X^n}$.
A distortion measure $d_n$ is defined as
$
d_n : {\cal X}^n \times {\cal Y}^n \rightarrow [0, +\infty).
$
An encoder 
$
f_n : {\cal X}^n \rightarrow \{ 0,1 \}^{\star}
$
is possibly stochastic and produces a non-prefix code.
A decoder 
$
g_n :  \{ 0,1 \}^{\star}  \rightarrow {\cal Y}^n
$
is deterministic.

We define an $(n, D, R, \epsilon, t)$ code as follows.
\begin{defi}
Given $n \in \mathbb{N}$, $D, R \geq 0$, $\epsilon \in [0,1)$, and $t > 0$, 
a code $(f_n, g_n)$ satisfying
\begin{align}
\mathbb{P} \left[ \frac{1}{n} d_n (X^n, g_n (f_n (X^n))) > D \right] &\leq \epsilon, \label{cd1} \\
\frac{1}{n t} \log \mathbb{E} [2^{t \ell(f_n (X^n))}]  &\leq  R \label{cd2}
\end{align}
is called an $(n, D, R, \epsilon, t)$ code.
\end{defi}

The fundamental limit is
\begin{align}
R^{*} (n, D, \epsilon, t) & := \inf \{ R : \mbox{$\exists$ {\rm an} $(n, D, R, \epsilon, t)$ {\rm code}} \}.
\end{align}

\section{Previous Study}
Courtade and Verd\'u \cite{Courtade2} considered the same problem setting with the restriction that
the code $(f, g)$ satisfies
$
\mathbb{P} [ d(X, g (f (X))) > D ]= 0
$
(i.e., $\epsilon = 0$ in (\ref{NP1})).
One of the main results in \cite{Courtade2} is
the converse bound on $R^{*} (D, 0, t)$.
Before describing the result,
we first introduce the {\it D-tilted information} \cite{Kostina12}
and {\it $D$-tilted R\'enyi entropy} \cite{Courtade2}.

Let $R(D)$ be the rate-distortion function, i.e.,
\begin{align}
R(D) = \min_{\substack{P_{Y |X } : \\
\mathbb{E} [d(X,Y)] \leq D}} I(X;Y),
\end{align}
where 
$I(X; Y)$ denotes the mutual information 
between random variables $X$ and $Y$, and 
$P_{Y|X}$ denotes a conditional probability distribution of $Y$ given $X$.
Assume that the minimum in the rate-distortion function $R(D)$ is achieved by $P^{\star}_{Y | X}$.
Further, let $Y^{\star}$ be a random variable taking a value in ${\cal Y}$ and whose distribution $P_{Y^{\star}}$ is the marginal of $P^{\star}_{Y | X} P_{X}$.
Then, the $D$-tilted information of $x \in {\cal X}$ is defined as\footnote{Kostina and Verd\'u\cite{Kostina12} named this quantity the $D$-tilted information. However, this quantity was used in earlier work by, e.g., Kontoyiannis \cite{Kontoyiannis00}.}
\begin{align}
\jmath_{X}(x,D) = \log\frac{1}{ \mathbb{E}[\exp \{  \lambda^{\star} D - \lambda^{\star} d(x, Y^{\star}) \}]},
\end{align}
where the expectation is with respect to $P_{Y^{\star}}$
and $\lambda^{\star} := - R' (D)$.
Further, the $D$-tilted R\'enyi entropy of order 
$\alpha \in (0,1) \cup (1,\infty)$ 
is defined as \cite{Courtade2}
\begin{align}
H_{\alpha} (X, D) = \frac{1}{1 - \alpha} \log \mathbb{E}[2^{(1-\alpha) \jmath_{X}(X,D)} ].
\end{align}

The next theorem characterizes the converse bound on $R^{*} (D, 0, t)$ by the $D$-tilted R\'enyi entropy.
\begin{theorem}[\cite{Courtade2}]
For any $D \geq 0$ and $t > 0$, 
\begin{align}
R^{*} (D, 0, t) \geq H_{\frac{1}{1+t}}(X, D) - \log \log (1+ \min\{|{\cal X}|, |{\cal Y}| \}),
\end{align}
where $|{\cal X}|$ and $|{\cal Y}|$ represent the cardinality of ${\cal X}$ and ${\cal Y}$, respectively.
\end{theorem}

\begin{remark}
The previous study \cite{Courtade2} investigated the case where the excess distortion probability is zero (i.e., $\epsilon = 0$ in (\ref{NP1})).
Further, they only showed the converse result.
On the other hand, our study deals with positive excess distortion probability as in (\ref{NP1}).
Moreover, our study investigates both achievability and converse bounds.
\end{remark}

\section{Non-asymptotic Analysis} \label{MainNonAsymptotic}

\subsection{Preliminary: R\'enyi Entropy-Based Quantity}
For $\alpha \in (0,1) \cup (1,\infty)$, the {\it R\'enyi entropy} is defined as \cite{Renyi}
\begin{align}
H_{\alpha}(X) = \frac{1}{1 - \alpha} \log \sum_{x \in {\cal X}} [P_{X}(x)]^{\alpha}.
\end{align}

One of the useful properties of the R\'enyi entropy
is {\it Schur concavity}.
This property is used in the proof of the achievability result in our main theorem.
To state the definition of a Schur concave function, we first review the notion of {\it majorization}.

\begin{defi}
Let $\mathbb{R}_{+}$ be the set of non-negative real numbers 
and $\mathbb{R}^{m}_{+}$ be the $m$-th Cartesian product of $\mathbb{R}_{+}$, where $m$ is a positive integer.
Suppose that ${\bf x} = (x_1, \ldots,$ $x_m) \in \mathbb{R}^{m}_{+}$ and ${\bf y} = (y_1, \ldots, y_m) \in \mathbb{R}^{m}_{+}$ satisfy
$
x_i \geq x_{i+1}$,  
$
y_i \geq y_{i+1}$
$
 (i=1,2, \ldots, m-1).
$
If ${\bf x} \in \mathbb{R}^{m}_{+}$ and ${\bf y} \in \mathbb{R}^{m}_{+}$ satisfy, for $k=1, \ldots, m-1$, 
$
\sum_{i=1}^{k} x_i \leq \sum_{i=1}^{k} y_i 
$
and
$
\sum_{i=1}^{m} x_i = \sum_{i=1}^{m} y_i ,
$
then we say that ${\bf y}$ {\it majorizes} ${\bf x}$ (it is denoted as ${\bf x} \prec {\bf y}$ in this paper).
\end{defi}

Schur concave functions are defined as follows.
\begin{defi}
We say that a function $h(\cdot): \mathbb{R}^{m}_{+} \rightarrow \mathbb{R}$ is a {\it Schur concave} function
if $h({\bf y}) \leq h({\bf x})$ for any ${\bf x}, {\bf y} \in \mathbb{R}^{m}_{+}$ satisfying ${\bf x} \prec {\bf y}$.
\end{defi}

For any $\alpha \in (0,1) \cup (1,\infty)$, 
the R\'enyi entropy $H_{\alpha}(X) $ is a Schur concave function (see, e.g., \cite{Marshall}).

Next, we introduce {\it a new quantity based on the R\'enyi entropy}.
This quantity plays an important role in producing our main results.

\begin{defi}
Given $D \geq 0$, $\epsilon \in [0,1)$, and $\alpha \in (0,1) \cup (1,\infty)$,
$G^{D, \epsilon}_{\alpha}(X)$ is defined as
\begin{align}
G^{D, \epsilon}_{\alpha}(X) = 
\min_{\substack{P_{Y|X} : \\
\mathbb{P} [ d(X, Y) > D ] \leq \epsilon}} H_{\alpha}(Y).
\end{align}
\end{defi}

\begin{remark}
For a given  $D \geq 0$ and $\epsilon \in [0,1)$, suppose that
\begin{align}
\mathbb{P} [ \inf_{y \in {\cal Y}} d(X, y) >D ] >\epsilon. \label{inf}
\end{align}
Then, there are no codes whose excess distortion probability is less than or equal to $\epsilon$.
Conversely, if such codes do not exist for given $D$ and $\epsilon$, (\ref{inf}) holds.
In this case, we define $R^{*} (D, \epsilon, t) = +\infty$.
Further, if (\ref{inf}) holds, we also define $G^{D, \epsilon}_{\alpha}(X) = +\infty$
because there is no conditional probability distribution $P_{Y|X}$ on ${\cal Y}$ satisfying
$\mathbb{P}[ d(X, Y) > D ] \leq \epsilon$.
\end{remark}

\subsection{Non-Asymptotic Coding Theorem}

The next lemma shows the achievability result on $R$ of a $(D, R, \epsilon, t)$ code.

\begin{lem} \label{OneShotAchievability}
For any $D \geq 0$, $\epsilon \in [0, 1)$, and $t > 0$,
there exists a $(D, R, \epsilon, t)$ code such that
\begin{align}
R=G^{D,\epsilon}_{\frac{1}{1+t}}(X) \label{3TH5}.
\end{align}
\end{lem}

\begin{IEEEproof}
See Section \ref{ProofOneShotAchievability}.
\end{IEEEproof}

\begin{remark}
The random coding argument is not used to prove the achievability result.
Instead, an explicit code construction is given.
This is similar to Feinstein's cookie-cutting argument \cite{Feinstein}.
\end{remark}

The next lemma shows the converse bound on $R$ of a $(D, R, \epsilon, t)$ code.

\begin{lem} \label{OneShotConverse}
For any $D \geq 0$, $\epsilon \in [0, 1)$, and $t > 0$, any $(D, R, \epsilon, t)$ code satisfies
\begin{align}
R \geq G^{D,\epsilon}_{\frac{1}{1+t}}(X) - \log \log (1+ \min\{|{\cal X}|, |{\cal Y}| \}). \label{oneshotconverse}
\end{align}
\end{lem}

\begin{IEEEproof}
See Section \ref{ProofOneShotConverse}.
\end{IEEEproof}

Combining Lemmas \ref{OneShotAchievability} 
and \ref{OneShotConverse}, 
we can immediately obtain the following result on $R^{*} (D, \epsilon, t)$.

\begin{theorem} \label{OneShotTheorem}
For any $D \geq 0$, $\epsilon \in [0, 1)$, and $t > 0$,
\begin{align}
G^{D,\epsilon}_{\frac{1}{1+t}}(X) - \log \log (1+ \min\{|{\cal X}|, |{\cal Y}| \}) 
& \leq R^{*} (D, \epsilon, t) \notag \\
& \leq G^{D,\epsilon}_{\frac{1}{1+t}}(X). \label{oneshotmain}
\end{align}
\end{theorem}

The same discussion 
which is used to prove Theorem \ref{OneShotTheorem} establishes the next result on 
$R^{*} (n, D, \epsilon, t)$.

\begin{theorem} \label{CodingTheorem}
For any $n \in \mathbb{N}$, $D \geq 0$, $\epsilon \in [0, 1)$, and $t > 0$,
\begin{align}
& \frac{1}{n} G^{D,\epsilon}_{\frac{1}{1+t}}(X^n) - \frac{1}{n} \log \log (1+ \min\{|{\cal X}^n|, |{\cal Y}^n| \}) \notag \\
& \leq R^{*} (n, D, \epsilon, t) 
\leq \frac{1}{n} G^{D,\epsilon}_{\frac{1}{1+t}}(X^n), \label{main}
\end{align}
where $G^{D,\epsilon}_{\frac{1}{1+t}}(X^n)$ is defined as 
\begin{align}
G^{D, \epsilon}_{\frac{1}{1+t}}(X^n) = 
\min_{\substack{P_{Y^n|X^n} : \\
\mathbb{P} [ d_n (X, Y) > n D ] \leq \epsilon}} H_{\frac{1}{1+t}}(Y^n).
\end{align}
\end{theorem}

\section{Asymptotic Analysis for a Stationary Memoryless Source}

This section investigates the general formula (\ref{main}) when a stationary memoryless source is assumed.
Especially, we consider the special case $t \downarrow 0$
and drive a single-letter characterization of the fundamental limit
$R^{*} (n, D, \epsilon, 0) := \lim_{t \downarrow 0} R^{*} (n, D, \epsilon, t)$.

First, two quantities are defined.
As we show in Section \ref{ProofThMemoryless}, 
they are closely related to the quantity 
$\lim_{\alpha \uparrow 1} G^{D,\epsilon}_{\alpha}(X^n)$.

\begin{defi}
Given $D \geq 0$ and $\epsilon \in [0,1)$, 
the {\it $(D, \epsilon)$-entropy} 
$H_{D, \epsilon}(X^n)$ is defined as \cite{Posner}
\begin{align}
H_{D, \epsilon}(X^n) = 
\min_{\substack{\varphi : {\cal X}^n \to {\cal Y}^n: \\
\mathbb{P} [ d_n (X^n, \varphi(X^n)) > n D ] \leq \epsilon}} H(\varphi(X^n)),
\end{align}
where $H(\cdot)$ denotes the Shannon entropy.
\end{defi}

\begin{defi}
Given $D \geq 0$ and $\epsilon \in [0,1)$, 
the quantity
$R_{D, \epsilon}(X^n)$ is defined as 
\begin{align}
R_{D, \epsilon} (X^n) = 
\min_{\substack{P_{Y^n | X^n} : \\
\mathbb{P} [ d_n (X^n, Y^n) > n D ] \leq \epsilon}} I(X^n; Y^n).
\end{align}
\end{defi}

Kostina et al.\ \cite{Kostina15} showed the next asymptotic result on $H_{D, \epsilon}(X^n)$ and $R_{D, \epsilon} (X^n)$.

\begin{theorem}[\cite{Kostina15}] \label{ThKostina15}
We impose the next assumptions:
\begin{itemize}
\item[$1)$]
For $(x^n, y^n) \in {\cal X}^n \times {\cal Y}^n$, 
the distortion measure $d_n(x^n, y^n)$ satisfies 
$
d_n(x^n, y^n) = \sum_{i=1}^{n} d(x_i, y_i). 
$

\item[$2)$] 
The distortion level $D$ satisfies $D \in (D_{\rm{min}}, D_{\rm{max}})$,
where $D_{\rm{min}} := \inf \{ D : R(D) < \infty \}$
and $D_{\rm{max}} := \inf_{y \in {\cal Y}} \mathbb{E} [d(X, y)]$.

\item[$3)$] 
The minimum in the rate-distortion function $R(D)$ is achieved by $P^{\star}_{Y | X}$.

\item[$4)$] 
$\mathbb{E} [d^{12} (X, Y^{\star})] < \infty$, where the expectation is with respect to $P_{X} \times P_{Y^{\star}}$.
\end{itemize}

Under a stationary memoryless source and the assumptions 1) -- 4), we have,
for any $\epsilon \in [0, 1)$, 
\begin{align}
& H_{D, \epsilon}(X^n) 
= R_{D, \epsilon} (X^n) \notag \\
& = (1- \epsilon) n R(D) - \sqrt{\frac{n V(D)}{2 \pi}} e^{-\frac{(Q^{-1}(\epsilon))^2}{2}}+ O(\log n),
\end{align}
where $V(D)$ is the {\it rate-dispersion function} \cite{Kostina12} which is defined as 
the variance of the $D$-tilted information, i.e.,
$
V (D) := {\rm Var} [\jmath_{X}(X,D)]
$
and 
$Q^{-1}(z)$ denotes the inverse function of 
$Q(z)= \int_{z}^{\infty} (1/\sqrt{2 \pi}) \exp(-t^{2}/2) dt$
for $z \in \mathbb{R}$.
\end{theorem}

Combination of Theorems \ref{CodingTheorem} and \ref{ThKostina15}
leads to the next single-letter characterization on 
$R^{*} (n, D, \epsilon, 0)$.
\begin{theorem} \label{ThMemoryless}
Under a stationary memoryless source and the assumptions 1) -- 4) in Theorem \ref{ThKostina15}, we have, for any $\epsilon \in [0, 1)$, 
\begin{align}
&\hspace{-3mm}  R^{*} (n, D, \epsilon, 0) \notag \\
&\hspace{-3mm} = (1- \epsilon) R(D) - \sqrt{\frac{V(D)}{2 \pi n}} e^{-\frac{(Q^{-1}(\epsilon))^2}{2}}+ O \left( \frac{\log n}{n} \right). \label{Memoryless}
\end{align}
\end{theorem}

\begin{IEEEproof}
See Section \ref{ProofThMemoryless}.
\end{IEEEproof}

\begin{remark} 
In view of Remark \ref{cumulant},
we observe that $R^{*} (n, D, \epsilon, 0)$ represents the fundamental limit of the mean codeword length.
This quantity was investigated by \cite{Kostina15}, and
our result (\ref{Memoryless}) coincides with the result in \cite{Kostina15}.
\end{remark}

\section{Proof of Main Results}

\subsection{Proof of Lemma \ref{OneShotAchievability}} \label{ProofOneShotAchievability}

First, some notations are defined before showing the construction of the encoder and the decoder.

\begin{itemize}
\item For any $y \in {\cal Y}$ and $D \geq 0$,
${\cal B}_D (y)$ is defined as
\begin{align}
{\cal B}_D (y) = \{ x \in {\cal X} : d(x,y) \leq D \}. \label{BD}
\end{align}

\item 
We define  $y_i $ ($i = 1, 2, \cdots$) by the following procedure.
Let $y_1$ be defined as
\begin{align}
y_1 = \argmax_{y \in {\cal Y}} \mathbb{P} [ X \in {\cal B}_{D} (y) ].
\end{align}
For $i = 2, 3, \cdots$, let $y_i $ be defined as
\begin{align}
y_i = \argmax_{y \in {\cal Y}} \mathbb{P} \left [ X \in {\cal B}_{D} (y) \setminus \bigcup_{j=1}^{i-1} {\cal B}_{D} (y_j) \right ].
\end{align}

\item For $i=1, 2, \ldots$, we define ${\cal A}_D(y_i)$ by
\begin{align}
{\cal A}_D(y_1) & = {\cal B}_{D} (y_1), \label{SetA1} \\
{\cal A}_D(y_i) & = {\cal B}_{D} (y_i) \setminus \bigcup_{j=1}^{i-1} {\cal B}_{D} (y_j) ~~ (\forall i \geq 2). \label{SetA}
\end{align}
From the definition, we have
\begin{align}
& \bigcup_{j=1}^{i} {\cal A}_{D} (y_j) = \bigcup_{j=1}^{i} {\cal B}_{D} (y_j) \quad (i \geq 1), \label{A1} \\
& {\cal A}_{D} (y_i) \cap {\cal A}_{D} (y_j) =  \emptyset \quad (\forall i \neq j), \label{A2} \\
& \mathbb{P} [ X \in {\cal A}_{D} (y_1) ] \geq
\mathbb{P} [ X \in {\cal A}_{D} (y_2) ] \geq \cdots. 
\end{align}

\item Given $\epsilon \in [0,1)$, 
let $k^{*} \geq 1$ be the integer satisfying 
\begin{align}
\sum_{i=1}^{k^{*}-1} \mathbb{P} [ X \in {\cal A}_{D} (y_i) ] &< 1- \epsilon, \label{ks1} \\
\sum_{i=1}^{k^{*}} \mathbb{P} [ X \in {\cal A}_{D} (y_i) ] & \geq 1- \epsilon. \label{ks2}
\end{align}

\item Let $\alpha$ and $\beta$ be defined as
\begin{align}
\alpha & = \sum_{i=1}^{k^{*}-1} \mathbb{P} [ X \in {\cal A}_{D} (y_i) ], \label{Alpha} \\
\beta & = 1-\epsilon-\alpha. \label{Beta}
\end{align}

\item Let $w_{i}$ be the $i$-th binary string in $\{ 0,1 \}^{\star}$ in the increasing order of the length and ties are arbitrarily broken.
For example, $w_1 = \lambda, w_2 = 0, w_3 = 1, w_4 = 00, w_5 = 01,$ etc.
\end{itemize}

Using these notations, we construct the following encoder 
$
\hat{f} : {\cal X} \rightarrow \{ 0,1 \}^{\star}
$
and decoder
$
\hat{g} :  \{ 0,1 \}^{\star}  \rightarrow {\cal Y}.
$

\medskip

\noindent
{\bf [Encoder]}
\begin{itemize}
\item[$ 1)$] For $x \in {\cal A}_{D} (y_i)$ ($i=1, \ldots, k^* -1$), set $\hat{f}(x) = w_i$.

\item[$2)$] For $x \in {\cal A}_{D} (y_{k^*})$,
set\footnote{
Note that we have 
$\mathbb{P} [ X \in {\cal A}_{D} (y_{k^*}) ] \geq \beta$ from (\ref{ks2}).
}
\begin{align}
\hspace{-12mm} \hat{f} (x) = \begin{cases}
    w_{k^*} &  {\rm with~ prob.} ~ \frac{\beta}{ \mathbb{P} [ X \in {\cal A}_{D} (y_{k^*}) ] }, \\
    w_1&  {\rm with~ prob.} ~ 1 - \frac{\beta}{ \mathbb{P} [ X \in {\cal A}_{D} (y_{k^*}) ] }.
  \end{cases}
\label{se}
\end{align}

\item[$3)$] For 
$x \notin  \bigcup_{i=1}^{k^*} {\cal A}_{D} (y_i)$,
set $\hat{f}(x) = w_1$.
\end{itemize}

\medskip

\noindent
{\bf [Decoder]}
Set $\hat{g}(w_i) = y_i$ ($i=1, \ldots, k^* $).
\medskip

Now, we evaluate the excess distortion probability.
We have
$d(x, \hat{g}(\hat{f}(x))) \leq D$ for $x \in {\cal A}_{D} (y_i)$
($i=1, \ldots, k^{*}-1$) since $\hat{g}(\hat{f}(x))=y_i$.
Furthermore, we have $d(x, \hat{g}(\hat{f}(x))) \leq D$ with probability 
$\beta/\mathbb{P} [ X \in {\cal A}_{D} (y_{k^*}) ]$ 
for
$x \in {\cal A}_{D} (y_{k^*})$. 
Thus,
\begin{align}
& \mathbb{P} [d(X, \hat{g} (\hat{f} (X))) \leq D ] \notag \\
& = \sum^{k^{*}-1}_{i=1} \mathbb{P} [ X \in {\cal A}_{D} (y_i) ] + \mathbb{P} [ \hat{f}(X) = w_{k^*}, X \in {\cal A}_{D} (y_{k^*}) ] \\
& =\alpha + \beta = 1 - \epsilon. 
\end{align}
Therefore, we have
$
\mathbb{P} [d(X, \hat{g} (\hat{f} (X))) > D ] = \epsilon.
$

Next, we evaluate the normalized cumulant generating function of codeword lengths for the code $(\hat{f}, \hat{g})$.
To this end, we denote by
$\hat{Y} := \hat{g} ( \hat{f}(X))$
and show the next lemma.

\begin{lem} \label{Alemma1}
For any $t > 0$ and $i \in \{1,2,\ldots,k^{*} \}$, 
we have
\begin{align}
2^{ t \ell(\hat{g}^{-1}(y_i))} \leq i^{t} \leq \left [ \sum_{j=1}^{k^*} \left (\frac{P_{\hat{Y}}(y_j)}{P_{\hat{Y}}(y_i)} \right )^{\frac{1}{1+t}} \right ]^{t}, \label{hodai}
\end{align}
where $\hat{g}^{-1}$ denotes the inverse function\footnote{From the construction of $\hat{g}$, we can define its inverse function.} of $\hat{g}$.
\end{lem}

\begin{IEEEproof}
First, we show the left inequality of (\ref{hodai}).
The construction of the code gives
\begin{align}
\ell(\hat{g}^{-1}(y_i)) \leq \log i
\end{align}
for any $i \in \{1,2,\ldots,k^{*} \}$.
This inequality yields
\begin{align}
2^{t \ell(\hat{g}^{-1}(y_i))} \leq 2^{t \log i} = i^{t},
\end{align}
which is the left inequality of (\ref{hodai}).

Next, we show the right inequality of (\ref{hodai}).
The code construction gives the next inequality
on the distribution of $\hat{Y}$:
\begin{align}
P_{\hat{Y}}(y_1) \geq P_{\hat{Y}}(y_2) \geq \ldots \geq P_{\hat{Y}}(y_{k^*}).
\end{align}
Thus, for any $i \in \{1,2,\ldots,k^{*} \}$, it follows that
\begin{align}
& \left (\frac{P_{\hat{Y}}(y_1)}{P_{\hat{Y}}(y_i)} \right )^{\frac{1}{1+t}} \geq 1, \quad
\left (\frac{P_{\hat{Y}}(y_2)}{P_{\hat{Y}}(y_i)} \right )^{\frac{1}{1+t}} \geq 1, \notag \\
& \ldots,
\left (\frac{P_{\hat{Y}}(y_{i-1})}{P_{\hat{Y}}(y_i)} \right )^{\frac{1}{1+t}} \geq 1, \quad
\left (\frac{P_{\hat{Y}}(y_{i})}{P_{\hat{Y}}(y_i)} \right )^{\frac{1}{1+t}} = 1. \label{ge1}
\end{align}
Hence, for any $i \in \{1,2,\ldots,k^{*} \}$, we have
\begin{align}
&i = \underbrace{1+1+\cdots+1}_{i} \\
&\overset{(a)}{\leq} \left (\frac{P_{\hat{Y}}(y_1)}{P_{\hat{Y}}(y_i)} \right )^{\frac{1}{1+t}} + \left (\frac{P_{\hat{Y}}(y_2)}{P_{\hat{Y}}(y_i)} \right )^{\frac{1}{1+t}} + \cdots + \left (\frac{P_{\hat{Y}}(y_{i})}{P_{\hat{Y}}(y_i)} \right )^{\frac{1}{1+t}}\\
& \overset{(b)}{\leq} \sum_{j=1}^{k^*} \left (\frac{P_{\hat{Y}}(y_j)}{P_{\hat{Y}}(y_i)} \right )^{\frac{1}{1+t}}, \label{right}
\end{align}
where $(a)$ follows from (\ref{ge1})
and
$(b)$ is due to
\begin{align}
\frac{P_{\hat{Y}}(y_j)}{ P_{\hat{Y}}(y_i)} \geq 0 \quad (\forall i, j \in \{1,2,\ldots,k^{*} \}).
\end{align}
The inequality (\ref{right}) yields the right inequality of (\ref{hodai}).
\end{IEEEproof}

Using Lemma \ref{Alemma1}, we have
\begin{align}
\mathbb{E} \left [ 2^{t \ell(\hat{g}^{-1}(\hat{Y}))} \right ] 
& = \sum_{i=1}^{k^*} P_{\hat{Y}}(y_i) 2^{t \ell(\hat{g}^{-1}(y_i))} \\
& \leq \sum_{i=1}^{k^*} P_{\hat{Y}}(y_i) \left [ \sum_{j=1}^{k^*} \left (\frac{P_{\hat{Y}}(y_j)}{P_{\hat{Y}}(y_i)} \right )^{\frac{1}{1+t}} \right ]^{t} \\
& = \left( \sum_{j=1}^{k^*} [P_{\hat{Y}}(y_j)]^{\frac{1}{1+t}} \right )^{1+t}. \label{mgf}
\end{align}
Thus, taking logarithm of both sides of (\ref{mgf}) and
dividing by $t > 0$, we have
\begin{align}
\frac{1}{t} \log \mathbb{E} \left [ 2^{t \ell(\hat{g}^{-1}(\hat{Y}))} \right ] 
& \leq \frac{1+t}{t} \log \sum_{j=1}^{k^*} [P_{\hat{Y}}(y_j)]^{\frac{1}{1+t}} \label{CumulantHat} \\
& = H_{\frac{1}{1+t}}(\hat{Y}). \label{cgf}
\end{align}

Finally, we evaluate the left and right-hand sides of
(\ref{cgf}).
The left-hand side of (\ref{cgf}) is evaluated as
\begin{align}
\mathbb{E} \left [ 2^{t \ell(\hat{g}^{-1}(\hat{Y}))} \right ]
= \mathbb{E} \left [ 2^{t \ell(\hat{f}(X))} \right ]. \label{cgfL}
\end{align}
Indeed, this is verified as follows:
\begin{align}
\mathbb{E} \left [ 2^{t \ell(\hat{f}(X))} \right ] 
&= \sum_{x \in {\cal X}} P_X(x) 2^{t \ell(\hat{f}(x))} \\
&= \sum_{i=1}^{k^*} \mathbb{P}[\hat{f}(X) = w_i] 2^{t \ell(w_i)} \\
&= \sum_{i=1}^{k^*} \mathbb{P}[\hat{g}^{-1}(\hat{Y}) = w_i] 2^{t \ell(w_i)} \\
&= \sum_{i=1}^{k^*} P_{\hat{Y}}(y_i) 2^{t \ell(\hat{g}^{-1}(y_i))} \\
&= \mathbb{E} \left [ 2^{t \ell(\hat{g}^{-1}(\hat{Y}))} \right ].
\end{align}
On the other hand, the right-hand side of (\ref{cgf}) is evaluated as
\begin{align}
H_{\frac{1}{1+t}}(\hat{Y}) = G^{D,\epsilon}_{\frac{1}{1+t}}(X). \label{cgfR}
\end{align}
This is proved by combining the fact that the R\'enyi entropy is a Schur concave function
and the next lemma shown in \cite{Saito}.
\begin{lem} [\cite{Saito}] \label{SAITO}
The distribution $P_{\hat Y}$ majorizes any $P_{\tilde{Y}}$ induced by $P_{\tilde{Y} |X}$ satisfying 
$\mathbb{P} [ d(X, \tilde{Y}) > D ] \leq \epsilon$.
\end{lem}

Therefore, the combination of (\ref{cgf}), (\ref{cgfL}), and (\ref{cgfR}) gives
\begin{align}
\frac{1}{t} \log \mathbb{E} \left [ 2^{t \ell(\hat{f}(X))} \right ] 
\leq G^{D,\epsilon}_{\frac{1}{1+t}}(X), \label{cgfG}
\end{align}
which completes the proof of Lemma \ref{OneShotAchievability}.

\subsection{Proof of Lemma \ref{OneShotConverse}} \label{ProofOneShotConverse}

Fix a $(D, R, \epsilon, t)$ code $(f, g)$ arbitrarily 
and we denote by $\overline{Y}:=g(f(X))$.
Further, without loss of generality, we assume that the decoder $g$ is an injective mapping\footnote{Note that it is sufficient to consider the case where the decoder $g$ is an injective mapping in the proof of the converse part (see, e.g., \cite{Courtade2}).}.
Then, the definition of a $(D, R, \epsilon, t)$ code gives
\begin{align}
\mathbb{P} [d(X, \overline{Y}) > D ] & \leq \epsilon, \label{katei1} \\
\frac{1}{t} \log \mathbb{E} [2^{t \ell(f(X))}] & \leq R, \label{katei2}
\end{align}
and the assumption that $g$ is an injective mapping yields the next inequality \cite{Courtade1}:
\begin{align}
\sum_{y \in \overline{{\cal Y}}} 2^{- \ell(g^{-1}(y))} 
& \leq \log (1+ \min\{|{\cal X}|, |{\cal Y}| \}), \label{C1}
\end{align}
where $\overline{{\cal Y}} := \{ g(f(x)) : x \in {\cal X} \} \subset {\cal Y}$.

The key lemma in the proof of the converse result 
is as follows.
\begin{lem} 
For any $t > 0$, we have
\begin{align}
\frac{1}{t} \log \mathbb{E} [2^{t \ell(g^{-1} (\overline{Y}))}]\geq H_{\frac{1}{1+t}}(\overline{Y}) - \log \sum_{y \in \overline{{\cal Y}}} 2^{- \ell(g^{-1}(y))}. \label{C2}
\end{align}
\end{lem}

\begin{IEEEproof}
For each $y \in \overline{{\cal Y}}$, 
\begin{align}
\alpha(y) & := \left[ 2^{\ell(g^{-1}(y))} \right ]^{-\frac{t}{1+t}} \label{alpha} \\
\beta(y) &:= \left[ P_{\overline{Y}} (y) \right ]^{\frac{1}{1+t}} \left[ 2^{\ell(g^{-1}(y))} \right ]^{\frac{t}{1+t}} \label{beta}.
\end{align}
Then, H\"{o}lder's inequality gives
\begin{align}
\sum_{y \in \overline{{\cal Y}}} \alpha(y) \beta(y)
\leq \left ( \sum_{y \in \overline{{\cal Y}}} [\alpha(y)]^{\frac{1+t}{t}} \right )^{\frac{t}{1+t}}\left ( \sum_{y \in \overline{{\cal Y}}} [\beta(y)]^{1+t} \right )^{\frac{1}{1+t}}. \label{Holder}
\end{align}
Taking logarithm of both sides of (\ref{Holder})
and substituting (\ref{alpha}) and (\ref{beta}) for (\ref{Holder}), we obtain
\begin{align}
& \frac{1+t}{t} \log \sum_{y \in \overline{{\cal Y}}} [P_{\overline{Y}}(y)]^{\frac{1}{1+t}} \notag \\ 
& \leq
\log \sum_{y \in \overline{{\cal Y}}} 2^{- \ell(g^{-1}(y))} + \frac{1}{t} \log \mathbb{E} [2^{t \ell(g^{-1}(\overline{Y})))}]. \label{CL}
\end{align}
Further, noticing that the left hand side of (\ref{CL}) is 
\begin{align}
\frac{1+t}{t} \log \sum_{y \in \overline{{\cal Y}}} [P_{\overline{Y}}(y)]^{\frac{1}{1+t}} = H_{\frac{1}{1+t}}(\overline{Y}),
\end{align}
we obtain the desired result (\ref{C2}).
\end{IEEEproof}

Combination of (\ref{C1}), (\ref{C2}), and
$
\mathbb{E} \left [ 2^{t \ell(f(X))} \right ] 
= \mathbb{E} \left [ 2^{t \ell(g^{-1}(\overline{Y}))} \right]
$
yields 
\begin{align}
&\hspace{-5mm} \frac{1}{t} \log \mathbb{E} \left [ 2^{t \ell(f(X))} \right ]  \notag \\
& \geq H_{\frac{1}{1+t}}(\overline{Y}) - \log \log (1+ \min\{|{\cal X}|, |{\cal Y}| \})]. \label{cp}
\end{align}
Finally, from (\ref{katei1}) and (\ref{katei2}),
we have (\ref{oneshotconverse}).

\subsection{Proof of Theorem \ref{ThMemoryless}} \label{ProofThMemoryless}

We denote by 
$
G^{D, \epsilon}_{1}(X^n)  := \lim_{\alpha \uparrow 1} G^{D, \epsilon}_{\alpha}(X^n)
$
and 
$\hat{Y}^n := \hat{g}_n ( \hat{f}_n (X^n))$,
where $(\hat{f}_n, \hat{g}_n)$ is the code as constructed in the proof of Lemma \ref{OneShotAchievability}.
Then, we have
\begin{align}
& G^{D, \epsilon}_{1}(X^n)  
= \lim_{\alpha \uparrow 1} G^{D, \epsilon}_{\alpha}(X^n) 
\overset{(a)}{=} \lim_{\alpha \uparrow 1} H_{\alpha}(\hat{Y}^n) \\
& \overset{(b)}{=} H (\hat{Y}^n) 
 \overset{(c)}{=} \min_{\substack{P_{Y^n|X^n} : \\
\mathbb{P} [ d_n (X^n, Y^n) > n D ] \leq \epsilon}} H (Y^n), \label{A1}
\end{align}
where
(a) follows from (\ref{cgfR}),
(b) is due to the fact that the R\'enyi entropy approaches the Shannon entropy as $\alpha$ tends to 1, and
(c) follows from Lemma \ref{SAITO} and the fact that the Shannon entropy is a Schur concave function (e.g., \cite{Marshall}).

Further, the definition of $H_{D, \epsilon}(X^n)$
gives
\begin{align}
\min_{\substack{P_{Y^n|X^n} : \\
\mathbb{P} [ d_n (X^n, Y^n) > n D ] \leq \epsilon}} H (Y^n)
\leq H_{D, \epsilon}(X^n). \label{A2}
\end{align}

Combination of (\ref{A1}) and (\ref{A2}) yields
\begin{align}
G^{D, \epsilon}_{1}(X^n)  \leq H_{D, \epsilon}(X^n). \label{A3}
\end{align}

On the other hand, we have
\begin{align}
& R_{D, \epsilon} (X^n) 
= \min_{\substack{P_{Y^n|X^n} : \\
\mathbb{P} [ d_n (X^n, Y^n) > n D ] \leq \epsilon}} I(X^n; Y^n) \\
& \overset{(a)}{\leq} \min_{\substack{P_{Y^n|X^n} : \\
\mathbb{P} [ d_n (X^n, Y^n) > n D ] \leq \epsilon}} H(Y^n) 
\overset{(b)}{=} G^{D, \epsilon}_{1}(X^n), \label{A4}
\end{align}
where 
(a) is due to the non-negativity of the conditional Shannon entropy and 
(b) follows from (\ref{A1}).

Thus, combination of (\ref{A3}) and (\ref{A4}) and application of Theorem \ref{ThKostina15} establish
\begin{align}
& G^{D, \epsilon}_{1}(X^n) \notag \\
& = (1- \epsilon) n R(D) - \sqrt{\frac{n V(D)}{2 \pi}} e^{-\frac{(Q^{-1}(\epsilon))^2}{2}}+ O(\log n). \label{A5}
\end{align}

Finally, 
letting $t \downarrow 0$ in Theorem \ref{CodingTheorem},
using (\ref{A5}),
and noticing
\begin{align}
\frac{1}{n} \log \log (1+ \min\{|{\cal X}^n|, |{\cal Y}^n| \}) = O \left (\frac{\log n}{n} \right),
\end{align} 
we obtain the desired result (\ref{Memoryless}).

\section{Discussion} \label{Discuss}

\subsection{Theorem for a Deterministic Code}
So far, we have treated a {\it stochastic code}.
If we deal with only a {\it deterministic code},
we have the next lemma instead of Lemma \ref{OneShotAchievability}.

\begin{lem} \label{Deterministic}
For any $D \geq 0$, $\epsilon \in [0, 1)$, define $\gamma$ as\footnote{Note that it holds that $\gamma \leq \epsilon$.} 
$\gamma = 1- \sum_{i=1}^{k^{*}} \mathbb{P} [ X \in {\cal A}_{D} (y_i)]$, 
where ${\cal A}_{D} (y_i)$ is defined as in (\ref{SetA1}) and (\ref{SetA})
and $k^*$ is the integer satisfying (\ref{ks1}) and (\ref{ks2}).
Then, for any $t > 0$,
there exists a deterministic $(D, R, \epsilon, t)$ code such that
\begin{align}
R=G^{D,\epsilon}_{\frac{1}{1+t}}(X) + \frac{(\epsilon - \gamma) \beta^{-\frac{t}{1+t} } \log e}{t \exp \left \{ \frac{t}{1+t} G^{D,\epsilon}_{\frac{1}{1+t}}(X) \right \} },
\end{align}
where $\beta$ is defined as in (\ref{Beta}). 
\end{lem}

\begin{IEEEproof}
See Appendix \ref{ProofTheoremDeterministic}.
\end{IEEEproof}

Comparing Lemmas \ref{OneShotAchievability} and 
\ref{Deterministic},
we observe that the result for the deterministic code is weaker than that of the stochastic code.
In the asymptotic regime, however, 
the restriction to only deterministic code is negligible since 
\begin{align}
\frac{(\epsilon - \gamma) \beta^{-\frac{t}{1+t}} \log e}{n t \exp \left \{ \frac{t}{1+t} G^{D,\epsilon}_{\frac{1}{1+t}}(X^n) \right \} } \to 0
\end{align}
holds as $n \to \infty$.

\subsection{Theorem for a Prefix Code}
We have discussed a code without the prefix constraints.
In this section, we discuss a result for an encoder
$
f^{\rm p} : {\cal X} \rightarrow \{ 0,1 \}^{\star}
$
and a decoder
$
g^{\rm p} :  \{ 0,1 \}^{\star}  \rightarrow {\cal Y}
$
when we assume that $f^{\rm p}$ produces a prefix code.

As shown in (\ref{Fundamental}),
we have defined $R^{*} (D, \epsilon, t)$ for a non-prefix code.
Similarly, we define
$R^{*}_{\rm p} (D, \epsilon, t)$
as the fundamental limit on the normalized cumulant generating function of codeword lengths for a prefix code $(f^{\rm p}, g^{\rm p})$.
Then, a modification of the proof of Lemmas \ref{OneShotAchievability} and \ref{OneShotConverse} yields the next result.

\begin{theorem} \label{OneShotTheoremPrefix}
For any $D \geq 0$, $\epsilon \in [0, 1)$, and $t > 0$,
\begin{align}
G^{D,\epsilon}_{\frac{1}{1+t}}(X) 
\leq R^{*}_{\rm p} (D, \epsilon, t) 
\leq G^{D,\epsilon}_{\frac{1}{1+t}}(X) +  \lfloor \log k^* \rfloor  + 1, \label{oneshotmainprefix}
\end{align}
where $k^*$ is the integer satisfying (\ref{ks1}) and (\ref{ks2}).
\end{theorem}

\begin{IEEEproof}
See Appendix \ref{ProofTheoremPrefix}.
\end{IEEEproof}

\subsection{Non-Asymptotics and Distortion Balls}

In our non-asymptotic analysis, 
the distortion $D$-ball around $y$ (i.e., (\ref{BD})) plays a crucial role.
On the other hand, in the previous studies of non-asymptotics for lossy compression \cite{Courtade2}, \cite{Kontoyiannis00}, \cite{Kontoyiannis02}, \cite{Kostina12}, \cite{Kostina15},
the distortion $D$-ball around $x$ (i.e., 
$
\tilde{{\cal B}}_D (x) := \{ y \in {\cal Y} : d(x,y) \leq D \}
$)
plays an important role.
Investigating the relation of approaches between previous works and our work is one of the future works.

\appendices
\section{Proof of Lemma \ref{Deterministic}} \label{ProofTheoremDeterministic}
Define $\gamma$ as 
$
\gamma = 1- \sum_{i=1}^{k^{*}} \mathbb{P} [ X \in {\cal A}_{D} (y_i)].
$
Note that it holds that $\gamma \leq \epsilon$.

Now, we construct the following {\it deterministic} encoder 
$
\hat{f}_{\rm det} : {\cal X} \rightarrow \{ 0,1 \}^{\star}
$
and decoder
$
\hat{g}_{\rm det} :  \{ 0,1 \}^{\star}  \rightarrow {\cal Y}.
$

\medskip

\noindent
{\bf [Encoder]}
\begin{itemize}
\item[$ 1)$] For $x \in {\cal A}_{D} (y_i)$ ($i=1, \ldots, k^*$), set $\hat{f}_{\rm det}(x) = w_i$.

\item[$2)$] For 
$x \notin  \bigcup_{i=1}^{k^*} {\cal A}_{D} (y_i)$,
set $\hat{f}_{\rm det}(x) = w_1$.
\end{itemize}

\medskip

\noindent
{\bf [Decoder]}
Set $\hat{g}_{\rm det}(w_i) = y_i$ ($i=1, \ldots, k^* $).
\medskip

First, we evaluate the excess distortion probability.
From the definition of the encoder and the decoder,
\begin{align}
\mathbb{P} [d(X, \hat{g}_{\rm det} (\hat{f}_{\rm det} (X))) \leq D ]  
&= \sum^{k^{*}}_{i=1} \mathbb{P} [ X \in {\cal A}_{D} (y_i) ] \\
& \geq  1 - \epsilon. 
\end{align}
Therefore, we have
\begin{align}
\mathbb{P} [d(X, \hat{g}_{\rm det} (\hat{f}_{\rm det} (X))) > D ] \leq \epsilon. 
\end{align}

Next, we evaluate the normalized cumulant generating function of codeword lengths for the code $(\hat{f}_{\rm det}, \hat{g}_{\rm det})$.
To this end, we denote by
$\hat{Y}_{\rm det} := \hat{g}_{\rm det} ( \hat{f}_{\rm det}(X))$.
For any $t > 0$, we have
\begin{align}
& \frac{1}{t} \log \mathbb{E} \left [ 2^{t \ell(\hat{g}^{-1}_{\rm det}(\hat{Y}_{\rm det}))} \right ] \notag \\
& \overset{(a)}{\leq} \frac{1+t}{t} \log \sum_{j=1}^{k^*} [P_{\hat{Y}_{\rm det}}(y_j)]^{\frac{1}{1+t}} \\
& = \frac{1+t}{t} \log \left ( [P_{\hat{Y}_{\rm det}}(y_1)]^{\frac{1}{1+t}} + \sum_{j=2}^{k^* - 1} [P_{\hat{Y}_{\rm det}}(y_j)]^{\frac{1}{1+t}} \right. \notag \\
& \hspace{42mm} \left. + [P_{\hat{Y}_{\rm det}}(y_{k^*})]^{\frac{1}{1+t}}\right ) \\
& \leq \frac{1+t}{t} \log \left ( [P_{\hat{Y}_{\rm det}}(y_1) + (\epsilon - \gamma)]^{\frac{1}{1+t}} \right. \notag \\
& \hspace{21mm} \left. + \sum_{j=2}^{k^* - 1} [P_{\hat{Y}_{\rm det}}(y_j)]^{\frac{1}{1+t}} \right. \notag \\
& \hspace{21mm} \left. + [P_{\hat{Y}_{\rm det}}(y_{k^*})- (\epsilon - \gamma)+ (\epsilon - \gamma)]^{\frac{1}{1+t}}\right ) \\
& \overset{(b)}{=} \frac{1+t}{t} \log \left ( [P_{\hat{Y}}(y_1)]^{\frac{1}{1+t}} + \sum_{j=2}^{k^* - 1} [P_{\hat{Y}}(y_j)]^{\frac{1}{1+t}} \right. \notag \\
& \hspace{21mm} \left. + [P_{\hat{Y}}(y_{k^*}) + (\epsilon - \gamma)]^{\frac{1}{1+t}}\right ) \\
& \overset{(c)}{\leq} \frac{1+t}{t} \log \left ( \sum_{j=1}^{k^*} [P_{\hat{Y}}(y_j)]^{\frac{1}{1+t}} + \frac{\epsilon - \gamma}{1+t} [P_{\hat{Y}}(y_{k^*})]^{- \frac{t}{1+t}} \right ) \\
& = \frac{1+t}{t} \log \left ( \sum_{j=1}^{k^*} [P_{\hat{Y}}(y_j)]^{\frac{1}{1+t}} + \frac{\epsilon - \gamma}{1+t} \beta^{- \frac{t}{1+t}} \right ) \\
& \overset{(d)}{\leq} \frac{1+t}{t} \left \{ \log \left ( \sum_{j=1}^{k^*} [P_{\hat{Y}}(y_j)]^{\frac{1}{1+t}} \right ) \right. \notag \\
& \hspace{21mm} \left.  + \frac{(\epsilon - \gamma)\beta^{-\frac{t}{1+t}} \log e}{(1+t)\sum_{j=1}^{k^*} [P_{\hat{Y}}(y_j)]^{\frac{1}{1+t}}} \right \} \\
& = \frac{1+t}{t} \log \left ( \sum_{j=1}^{k^*} [P_{\hat{Y}}(y_j)]^{\frac{1}{1+t}} \right )
+ \frac{(\epsilon - \gamma) \beta^{-\frac{t}{1+t}} \log e}{t \sum_{j=1}^{k^*} [P_{\hat{Y}}(y_j)]^{\frac{1}{1+t}}}  \\
& \overset{(e)}{=} G^{D,\epsilon}_{\frac{1}{1+t}}(X) + \frac{(\epsilon - \gamma) \beta^{-\frac{t}{1+t}} \log e}{t \exp \left \{ \frac{t}{1+t} G^{D,\epsilon}_{\frac{1}{1+t}}(X) \right \} },
\end{align}
where
$(a)$ follows from the same discussion as in (\ref{CumulantHat}),
$(b)$ is due to the construction of $(\hat{f}, \hat{g})$ and $(\hat{f}_{\rm det}, \hat{g}_{\rm det})$,
$(c)$ and $(d)$ follow from Taylor's expansion,
and
$(e)$ is due to (\ref{cgfR}).

Thus, we complete the proof of Lemma \ref{Deterministic}.

\section{Proof of Theorem \ref{OneShotTheoremPrefix}} \label{ProofTheoremPrefix}

We define a $(D, R, \epsilon, t)_{\rm p}$ code as follows.
\begin{defi}
Given $D, R \geq 0$, $\epsilon \in [0,1)$, and $t > 0$, 
a prefix code $(f^{\rm p}, g^{\rm p})$ satisfying
\begin{align}
\mathbb{P} [ d(X, g^{\rm p} (f^{\rm p} (X))) > D ] &\leq \epsilon, \\
\frac{1}{t} \log \mathbb{E} [2^{t \ell(f^{\rm p}(X))}]  &\leq R 
\end{align}
is called a $(D, R, \epsilon, t)_{\rm p}$ code.
\end{defi}

Then, the fundamental limit that we investigate is
\begin{align}
R^{*}_{\rm p} (D, \epsilon, t) := \inf \{ R : \mbox{$\exists$ {\rm a} $(D, R, \epsilon, t)_{\rm p}$ {\rm code}} \}. 
\end{align}

To show Theorem \ref{OneShotTheoremPrefix},
we prove the next two lemmas.
If we prove these lemmas,
we can immediately obtain Theorem \ref{OneShotTheoremPrefix}.

\begin{lem} \label{OneShotAchievabilityPrefix}
For any $D \geq 0$, $\epsilon \in [0, 1)$, and $t > 0$,
there exists a $(D, R, \epsilon, t)_{\rm p}$ code such that
\begin{align}
R=G^{D,\epsilon}_{\frac{1}{1+t}}(X) +  \lfloor \log k^* \rfloor  + 1,
\end{align}
where $k^*$ is the integer satisfying (\ref{ks1}) and (\ref{ks2}).
\end{lem}

\begin{lem} \label{OneShotConversePrefix}
For any $D \geq 0$, $\epsilon \in [0, 1)$, and $t > 0$, any $(D, R, \epsilon, t)_{\rm p}$ code satisfies
\begin{align}
R \geq G^{D,\epsilon}_{\frac{1}{1+t}}(X).
\end{align}
\end{lem}

\medskip

{\it Proof of Lemma \ref{OneShotAchievabilityPrefix}:}
We use the same notations defined in Section \ref{ProofOneShotAchievability}.
Further, we introduce the next notation:
for $i=1, \ldots, k^*$,
a codeword $w^{\rm p}_{i}$ is defined as
\begin{align}
w^{\rm p}_{i} = w_i \circ h_i,
\end{align}
where $\circ$ denotes a concatenation and 
$h_i \in \{ 0,1 \}^{\star}$ is a binary sequence such that 
$w^{\rm p}_{i} \neq w^{\rm p}_{j} ~ (\forall i \neq j)$
and 
$|| h_i || = \lfloor \log k^* \rfloor  + 1 - \lfloor \log i \rfloor$,
where $|| \cdot ||$ denotes a length of a codeword.
The definition of $w^{\rm p}_{i}$ indicates that 
the length of $w^{\rm p}_{i}$ is $\lfloor \log k^* \rfloor  + 1$ for all $i=1, \ldots, k^*$.\footnote{Note that the length of $w_i$ is $\lfloor \log i \rfloor$.}

\begin{figure} [t]
\centering
\includegraphics[width=9.5cm,clip]{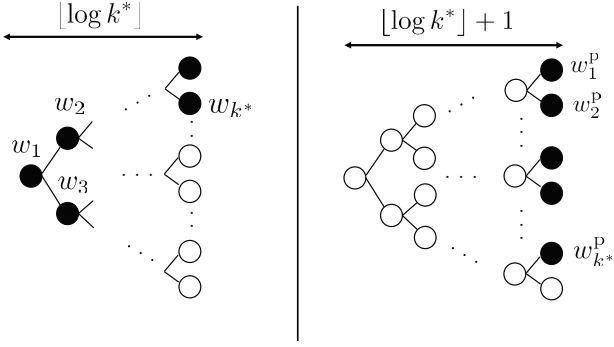}
\caption{Illustration of $w_1, w_2, \ldots, w_{k^*}$ and $w^{\rm p}_{1}, w^{\rm p}_{2}, \ldots, w^{\rm p}_{k^*}$ in a code tree.}
\label{codetree}
\end{figure}

Since the number of codewords 
$w_1, w_2, \ldots, w_{k^*}$ is at most
\begin{align}
1 + 2 + 2^2 + \cdots + 2^{\lfloor \log k^* \rfloor}
= 2^{\lfloor \log k^* \rfloor + 1} -1,
\end{align}
the codewords $w^{\rm p}_{1}, w^{\rm p}_{2}, \ldots, w^{\rm p}_{k^*}$ correspond to the leaf nodes of a code tree whose depth is $\lfloor \log k^* \rfloor + 1$
(see Fig.\ \ref{codetree}).
Thus, we can construct the following prefix code
$
\hat{f}^{\rm p} : {\cal X} \rightarrow \{ 0,1 \}^{\star}
$
and 
$
\hat{g}^{\rm p} :  \{ 0,1 \}^{\star}  \rightarrow {\cal Y}.
$

\medskip

\noindent
{\bf [Encoder]}
\begin{itemize}
\item[$ 1)$] For $x \in {\cal A}_{D} (y_i)$ ($i=1, \ldots, k^* -1$), set $\hat{f}^{\rm p}(x) = w^{\rm p}_{i}$.

\item[$2)$] For $x \in {\cal A}_{D} (y_{k^*})$,
set
\begin{align}
\hspace{-12mm} \hat{f}^{\rm p} (x) = \begin{cases}
    w^{\rm p}_{k^*} &  {\rm with~ prob.} ~ \frac{\beta}{ \mathbb{P} [ X \in {\cal A}_{D} (y_{k^*}) ] }, \\
    w^{\rm p}_1&  {\rm with~ prob.} ~ 1 - \frac{\beta}{ \mathbb{P} [ X \in {\cal A}_{D} (y_{k^*}) ] }.
  \end{cases}
\end{align}

\item[$3)$] For 
$x \notin  \bigcup_{i=1}^{k^*} {\cal A}_{D} (y_i)$,
set $\hat{f}^{\rm p}(x) = w^{\rm p}_{1}$.
\end{itemize}

\medskip

\noindent
{\bf [Decoder]}
Set $\hat{g}^{\rm p}(w_i) = y_i$ ($i=1, \ldots, k^* $).
\medskip

Now, we evaluate the excess distortion probability of the code $(\hat{f}^{\rm p}, \hat{g}^{\rm p})$.
The same discussion as in the proof of Lemma \ref{OneShotAchievability} yields
\begin{align}
\mathbb{P} [d(X, \hat{g}^{\rm p} (\hat{f}^{\rm p} (X))) > D ] = \epsilon. 
\end{align}

Next, we evaluate the normalized cumulant generating function of codeword lengths for the code $(\hat{f}^{\rm p}, \hat{g}^{\rm p})$:
\begin{align}
& \frac{1}{t} \log \mathbb{E} \left [ 2^{t \ell(\hat{f}^{p}(X))} \right ] \\
& \overset{(a)}{\leq} \frac{1}{t}  \log \sum_{x \in {\cal X}} P_X (x) 2^{t \ell(\hat{f}(x)) + t (\lfloor \log k^* \rfloor  + 1)} \\
& = \frac{1}{t}  \log \left (\sum_{x \in {\cal X}} P_X (x) 2^{t \ell(\hat{f}(x))} \right ) + \lfloor \log k^* \rfloor  + 1 \\
& = \frac{1}{t} \log \left ( \mathbb{E} \left [ 2^{t \ell(\hat{f}(X))} \right ] \right ) + \lfloor \log k^* \rfloor  + 1 \\
& \overset{(b)}{\leq} G^{D,\epsilon}_{\frac{1}{1+t}}(X) + \lfloor \log k^* \rfloor  + 1,
\end{align}
where 
$(a)$ follows from the construction of 
$(\hat{f}, \hat{g})$ in Section \ref{ProofOneShotAchievability} and that of $(\hat{f}^{\rm p}, \hat{g}^{\rm p})$,
and 
$(b)$ is due to (\ref{cgfG}).
This completes the proof of Lemma \ref{OneShotAchievabilityPrefix}.

\medskip

{\it Proof of Lemma \ref{OneShotConversePrefix}:}

By replacing (\ref{C1}) with Kraft's inequality and 
following the same route as in the proof of Lemma \ref{OneShotConverse},
we obtain Lemma \ref{OneShotConversePrefix}.


\section*{Acknowledgment}
The authors would like to thank Dr.\ Hideki Yagi for helpful discussions.
This work was supported in part by JSPS KAKENHI Grant Numbers 
JP16K00195,
JP16K00417,
JP17K00316,
JP17K06446,
and by Waseda University Grant for Special Research Projects (Project number: 2017A-022).




\end{document}